\documentclass{iopart}

\usepackage{graphics}

\begin{document}

\title{Maximum Fidelity Retransmission of Mirror Symmetric Qubit States}
\author{Kieran Hunter$^1$, Erika Andersson$^1$, Claire R. Gilson$^2$,
and Stephen M. Barnett$^1$}
\address{$^1$Department of Physics, University of
Strathclyde, Glasgow G4 0NG, Scotland}
\address{$^2$ Department of Mathematics, University of Glasgow, Glasgow
G12 8QW, Scotland}

\begin{abstract}

In this paper we address the problem of optimal reconstruction of a quantum state
from the result of a single measurement
when the original quantum state is known to be a member of some
specified
set. A suitable figure of merit for this process is the fidelity, which is the
probability that the state we construct on the basis of the measurement result
is found
by a subsequent test to match the original state. We consider the maximisation of
the fidelity for a set of three mirror symmetric qubit states.
In contrast to previous examples, we find that the strategy which minimises the
probability of erroneously identifying the state does not generally maximise the
fidelity.

\end{abstract}


\section{Introduction}

The principles governing communication through a quantum channel have been
extensively studied. 
The transmitting agent, conventionally called Alice, selects a state from a
predefined set \{$|\psi_{j}\rangle$\} with relative frequency $p_{j}$ and
transmits a quantum system prepared in
this state through the quantum channel. The classical information that is
Alice's message is encoded on a string of such states. The receiving agent,
conventionally called Bob, knows the set of states \{$|\psi_{j}\rangle$\} and
their
relative frequencies $p_{j}$ which are the prior probabilities he assigns to the
states before he makes a measurement. Bob must make a measurement on the states
he receives to attempt to recover the encoded information.

The problem for Bob is to determine the best measurement to make. Which
measurement is best will depend on how the results are to be used, that is how the
information was encoded or what question about the signal states the measurement
is designed to answer. For each such coding or question  one can define a
mathematical `figure of merit' function which provides a measure of how appropriate
a given measurement strategy is. Bob's task in finding his optimal measurement
is to maximise this function with respect to all possible measurements. Commonly
considered examples of this are the minimum error probability (or minimum Bayes
cost) \cite{Helstrom, Holevo, Yuen, Clarke1}
and the accessible information \cite{Clarke1, Davies, Sasaki, Mizuno},
both of which describe recovery of classical information about the original
message.

For some applications, Bob needs to use his measurement result to reproduce the
quantum signal. The
objective is then that the new signal matches the original as closely as
possible. We must now consider optimal strategies for the combined measurement
and reconstruction process. The quality of these measurement-retransmission
strategies is associated with the {\em fidelity} $F$.
This is the probability that a
subsequent measurement on the
retransmitted state will confirm that it matches the original signal. The
fidelity is
 a measure of the recovery of quantum information about the state of the
signal rather than classical information about the original message.

One motivation for this discussion of the fidelity is the well known question of
eavesdropping in quantum communication. Here a third hostile party
(conventionally called Eve) wishes to measure the signal coming from Alice and
transmit a new signal to Bob which matches the original as closely as possible.
Then the error probability or the accessible information of the measurement
strategy describes how much Eve expects to gain from eavesdropping, but the
fidelity determines the probability that her interception of the signal is
undetected. In cryptographic key distribution problems Eve has additional options such as an
imperfect cloning and delayed measurement strategy, to take advantage of the
extra information available when the preparation basis is announced. It is
obviously more appropriate to consider the cloning fidelity \cite{Gisin, Bruss98} than 
the
retransmission fidelity for such strategies.

No general condition is known for maximising the
fidelity of a measurement and retransmission strategy, 
but the maximum fidelity has been found for specific cases
\cite{Bfid,F2Gen}. These cases are when the possible signals form a set of
symmetric qubit states \cite{Bfid}, and where there are only two possible signal
qubit states \cite{F2Gen}. For these cases the measurement strategy which
minimises the probability of incorrectly identifying the states always maximises
the fidelity for the best choice of retransmission states. This optimal
strategy
is not, however, unique for sets of three or more symmetric states. 

It is interesting to ask whether the strategy that minimises the error probability
always maximises the fidelity. If this is the case then our best strategy is to identify
the original signal state as well as we can and then select a corresponding
retransmission state. In this paper we establish  that the fidelity is not always
maximised by the strategy which minimises the probability of erroneously identifying the
signal state. We demonstrate this by maximising the fidelity for the mirror-symmetric
qubit states for which the minimum error strategy has recently been derived
\cite{Erika1}.

\section{Fidelity}

The previous work on maximum fidelity for symmetric states \cite{Bfid}
established some
important results which we shall make use of. We shall use the notation
contained in
that work. The signal states are denoted by $|\psi_{j}\rangle$ with
associated
prior probabilities $p_{j}$ and the retransmission states are $|\phi_{k}\rangle$.
 The
measurement is described by its Probability Operator Measure (POM) 
 elements $\hat{\Pi}_{k}$. These POM elements
 are operators which represent the probability of occurrence of each possible
 outcome of a measurement. The probability $P(k|j)$ of the outcome $k$ occurring
 given that the system was prepared in the state $|\psi_{j}\rangle$ is
 \begin{equation}
P(k|j) = \langle\psi_{j}|\hat{\Pi}_{k}|\psi_{j}\rangle.
\end{equation}

For the POM elements  $\hat{\Pi}_{k}$ to represent probabilities, they must be
subject to
the following constraints:
\begin{enumerate}
\item All the $\hat{\Pi}_{k}$'s are Hermitian.
\item Their eigenvalues are non-negative.
\item The total probability of all outcomes for any input sums to 1:
\begin{equation}
\sum_{k} \hat{\Pi}_{k} = \hat{1}.\label{eq:POMcond2}
\end{equation}
\end{enumerate}

To find the optimal measurement-retransmission strategy we need to express the
fidelity $F$ as a function of the POM elements $\hat{\Pi}_{k}$, the
retransmission states $|\phi_{k}\rangle $ and
the set of possible signal states $\{|\psi_{j}\rangle, p_{j}\}$.
The
fidelity is the
probability that the state $|\phi_{k}\rangle$ selected on the
basis of the measurement outcome $\hat{\Pi}_{k}$  will pass a test of the
question
`Is the state $|\psi_{i}\rangle$?'. This test is described by the POM
\mbox{ \{ $|\psi_{i}\rangle \langle \psi_{i}|,
\hat{1} - |\psi_{i}\rangle \langle \psi_{i}|$\},} where $|\psi_{i}\rangle$ is
the state of the
original signal. Thus $F$ is given by
\begin{equation}
F = \sum_{j,k} |\langle \psi_{j}|\phi_{k}\rangle|^2 \langle
\psi_{j}|\hat{\Pi}_{k}|\psi_{j}\rangle p_{j}.
\end{equation}
This can be written as \cite{ Bfid, F2Gen}
\begin{equation}
F = \sum_{k} \langle \phi_{k}|\hat{O}_{k}|\phi_{k} \rangle, \label{eigen}
\end{equation}
where the positive operator $\hat{O}_{k}$ is given by
\begin{equation}
\hat{O}_{k} = \sum_{j} |\psi_{j}\rangle \langle
\psi_{j}|\hat{\Pi}_{k}|\psi_{j}\rangle \langle \psi_{j}|p_{j}. \label{Ok}
\end{equation}
It is clear from this that the optimal retransmission states
$|\phi_{k}\rangle$ are the eigenvectors of
the operators $\hat{O}_{k}$ corresponding to the largest eigenvalue $\nu_{k_{+}}$ of
$\hat{O}_{k}$. 

If these optimal retransmission states are used, then the fidelity is given by
the sum of the largest eigenvalues of the
$\hat{O}_{k}$ operators:
\begin{equation}
F = \sum_{k} \nu_{k_{+}}
\end{equation}
and we need only consider the maximisation of the largest eigenvalues of
$\hat{O}_{k}$, subject to the constraint that the operators $\hat{\Pi}_{k}$
form a POM. In such a maximisation each of the POM elements $\hat{\Pi}_{k}$ can be
assumed to be proportional to a pure state projector, since the action of any
mixed-state like
element here would be identical to that of a number of rank 1 POM elements
corresponding to the same retransmission state.

\section{Mirror Symmetric States}

We have recently described the minimum error
strategy for a qubit which
 is
known to be one of a set of three mirror symmetric qubit states \cite{Erika1}. Here mirror
symmetric means that the set of states
 is invariant under the transformation
\begin{equation}
|+\rangle \longrightarrow +|+\rangle , \  |-\rangle \longrightarrow -|-\rangle,
\end{equation}
with the prior probabilities associated with any mirror-symmetric pair of states
being equal.

The set of three mirror symmetric qubit states can be written as
\begin{eqnarray}
|\psi_{1}\rangle & = & \cos{\theta}|+\rangle + \sin{\theta}|-\rangle, \ p_{1} =
p, \nonumber \\
|\psi_{2}\rangle & = & \cos{\theta}|+\rangle - \sin{\theta}|-\rangle, \ p_{2} =
p, \nonumber \\
|\psi_{3}\rangle & = & |+\rangle, \ p_{3} = 1-2p \label{psi};
\end{eqnarray}
where $0 \leq \theta \leq \frac{\pi}{2}$ and $0 \leq p \leq \frac{1}{2}$.

The minimum error strategy was found to be of different form in two distinct
domains of $p$ and $\theta$. The solutions in
these two domains are: \\
for
\begin{equation}
p \geq \frac{1}{2+\cos{\theta} (\cos{\theta}+\sin{\theta})} \label{pe2opt}
\end{equation}
the minimum error measurement strategy is given by
\begin{eqnarray}
\hat{\Pi}_{1}  & = & \frac{1}{2} (|+\rangle +|-\rangle )(\langle +|+ \langle -|),
\nonumber  \\
\hat{\Pi}_{2}  & = & \frac{1}{2} (|+\rangle -|-\rangle )(\langle +|-\langle -|),
\nonumber \\
\hat{\Pi}_{3}  & = & 0; \label{2elements}
\end{eqnarray}
and for
\begin{equation}
p \leq \frac{1}{2+\cos{\theta} (\cos{\theta}+\sin{\theta})}
\end{equation}
the minimum error measurement strategy is given by
\begin{eqnarray}
\hat{\Pi}_{1} & = & \frac{1}{2} (a|+\rangle +|-\rangle )(a\langle +|+\langle -|),
\nonumber  \\
\hat{\Pi}_{2} & = & \frac{1}{2} (a|+\rangle -|-\rangle )(a\langle +|-\langle -|),
\nonumber  \\
\hat{\Pi}_{3} & = & (1-a^2)|+\rangle \langle +|, \label{3elements}
\end{eqnarray}
where $a$ is the following function of $p$ and $\theta$:
\begin{equation}
a = \frac{p\cos{\theta}\sin{\theta}}{1-p(2+{\cos}^2{\theta})}.
\end{equation}

At the boundary between the two
domains, which is when the equality holds in the condition (\ref{pe2opt}),
$a=1$ and thus $\hat{\Pi}_{3} = 0$.

\section{Maximising Fidelity for the Mirror Symmetric States}

To find the maximum fidelity for the mirror symmetric states we will follow a
similar method to that used for the symmetric states \cite{Bfid}. We attempt
to
write an explicit formula for the fidelity in terms of some parameter set and
find the maximum by differentiation.

To maximise the fidelity for these mirror symmetric states we choose a
representation of the operator $\hat{O}_{k}$ and find its eigenvalues. To do
this we
 first obtain a general representation of the qubit POM elements. As we
stated
in section 2 we need only consider elements of rank 1, that is elements
proportional to pure state projectors.

The elements of such a POM can be represented by the matrices
\begin{equation}
\hat{\Pi}_{k} = w_{k} \left( \begin{array}{cc}
1+\cos{\theta_{k}} & \sin{\theta_{k}}e^{i\phi_{k}} \\
\sin{\theta_{k}}e^{-i\phi_{k}} & 1-\cos{\theta_{k}}
\end{array} \right);
\end{equation}
where the basis vectors are
\begin{equation}
|+\rangle = \left( \begin{array}{c}
1 \\
0
\end{array} \right),
|-\rangle = \left( \begin{array}{c}
0 \\
1
\end{array} \right);
\end{equation}
and $0 \leq w_{k} \leq \frac{1}{2}$, $-\pi < \theta_{k} \leq \pi$, 
$-\frac{\pi}{2} < \phi_{k} \leq \frac{\pi}{2}$.

These POM elements are automatically Hermitian and positive. The remaining
completeness constraint (\ref{eq:POMcond2}) becomes equations for $w_{k}$,
$\theta_{k}$ and $\phi_{k}$:
\begin{equation}
1 - \sum_{k} w_{k}  =  0, \label{con1} \end{equation}
\begin{equation}
\sum_{k} w_{k} \cos{\theta_{k}}  =  0, \label{con2} \end{equation}
\begin{equation}
\sum_{k} w_{k}  \sin{\theta_{k}}e^{i\phi_{k}}  =  0. \label{con3}
\end{equation}
The operators $\hat{O}_{k}$ for the set of three mirror symmetric states become
\begin{equation}
\fl \hat{O}_{k} =  w_{k} \left( \begin{array}{cc}
\begin{array}{c} { 
2p{\cos}^{2}{\theta}(1+\cos{2\theta}\cos{\theta_{k}})} \\ { 
+
(1-2p)(1+\cos{\theta_{k}}) } \end{array} &
{ 
p{\sin}^{2}{2\theta}\sin{\theta_{k}}\cos{\phi_{k}} } \\
  & \\
 { 
 p{\sin}^{2}{2\theta}\sin{\theta_{k}}\cos{\phi_{k}} }&
{ 
2p{\sin}^{2}{\theta}(1+\cos{2\theta}\cos{\theta_{k}}) }
\end{array} \right). \label{odef}
\end{equation}
The eigenvalues $\nu_{k_{\pm}}$ of this matrix are given by
\begin{equation}
 \begin{array}{lcl}
\frac{\nu_{k_{\pm}}}{w_{k}} & = &
 { 
 p(1+\cos{2\theta}\cos{\theta_{k}})+
(\frac{1}{2}-p)(1+\cos{\theta_{k}})} \\  & & { 
\pm
[(p\cos{2\theta} (1+\cos{2\theta}\cos{\theta_{k}}) +
(\frac{1}{2}-p)(1+\cos{\theta_{k}}))^{2} }  \\ & &  { 
+ p^{2}
{\sin}^{4}{2\theta}{\sin}^{2}{\theta_{k}}{\cos}^{2}{\phi_{k}}}]^{\frac{1}{2}},
\end{array} \label{ni}
\end{equation}
of which the greater eigenvalue is clearly $\nu_{k_{+}}$.

From the form of these eigenvalues we see that the fidelity $F$ is not a function
of $\sin{\theta_{k}}$. This means that any element with the parameters $(w_{k}, \theta_{k},
\phi_{k})$ gives the same contribution to the fidelity as would an element with the
parameters $(w_{k}, -\theta_{k}, \phi_{k})$, and thus the same contribution as the
pair of elements $(\frac{w_{k}}{2}, \theta_{k}, \phi_{k})$ and $(\frac{w_{k}}{2},
-\theta_{k}, \phi_{k})$. Thus we can replace all of the elements in a POM with such pairs
of elements without changing the fidelity, and we need only find the maximum fidelity for
POMs consisting of such pairs. Such POMs satisfy condition (\ref{con3}) automatically.
Since there is now no condition restricting our choice of $\phi_{k}$, we are free to select
$\phi_{k}$ to maximise each eigenvalue $\nu_{k_{+}}$ independently. It is clear from examination
of (\ref{ni}) that the best choice is always $\phi_{k}=0$ and thus $\cos{\phi_{k}}=1$.
In truth we should have expected such a symmetry of our measurement
strategy,
since this simply corresponds to the POM also being both mirror symmetric and
confined to the plane of the states \{$|\psi_{j}\rangle$\}.

Since the pair of elements corresponding to $\pm \theta_{k}$ are equally
weighted and each contributes the same amount to the fidelity, we now use the
parameter $w_{k}$
to represent the combined weight of the pair of elements with the same value
of
$\cos{\theta_{k}}$.

We can then write the eigenvalues as
\begin{equation}
\nu_{k_{+}} = w_{k} \left[ { 
p \left(1+\cos{2\theta}\cos{\theta_{k}}\right)+
\left(\frac{1}{2}-p\right) \left(1+\cos{\theta_{k}}\right) +
Q_{k}^{\frac{1}{2}}} \right],
\end{equation}
where the functions $Q_{k}$ are given by
\begin{equation}
\begin{array}{lcl}
Q_{k} & = & \left[p\cos{2\theta} \left(1+\cos{2\theta}\cos{\theta_{k}}\right) +
\left(\frac{1}{2}-p\right)\left(1+\cos{\theta_{k}}\right)\right]^{2} \\ & & +
p^{2}\sin^{4}{2\theta}\sin^{2}{\theta_{k}}.
\end{array}
\end{equation}
The POM constraints (\ref{con1}, \ref{con2}) allow us to simplify the
fidelity $F$:
\begin{equation}
F = \sum_{k} \nu_{k_{+}} = \frac{1}{2} + \sum_{k} w_{k} Q_{k}^{\frac{1}{2}}.
\label{qfid}
\end{equation}

To find the stationary points of $F$, subject to the constraints (\ref{con1},
\ref{con2}) on $w_{k}$ and
$\theta_{k}$ we shall use Lagrange's method of undetermined multipliers. We can
 construct the function $G$:
\begin{equation}
G = F + \alpha_{1} \left(1 - \sum_{k} w_{k}\right) + \alpha_{2} \left(\sum_{k} w_{k}
\cos{\theta_{k}}\right),
\end{equation}
with the constraint (\ref{con3}) being irrelevant since this $F$ depends on neither
$\sin{\theta_{k}}$ nor $\phi_{k}$. The full detail of the maximisation
calculation can be found in appendix A, but the main points are summarised here.

The equation $\frac{\partial G}{\partial \theta_{k}} = 0$ has four solutions for
$\theta_{k}$: $\theta_{k} = 0, \pi$ and $\pm \Omega$, where $ \Omega$ is some
angle depending on the unknown multiplier $\alpha_{2}$. This limits the number
of elements in any optimal
POM to four.

The minimum value of $w_{k}$, $w_{k} = 0$, clearly corresponds to trivial zero
operators. The maximum value of $w_{k}$ for a mirror symmetric pair of elements
arises from the positivity of the POM
elements and the completeness condition (\ref{eq:POMcond2}). These conditions
impose a tight bound on 
$\langle +|\hat{\Pi}_{k}|+\rangle$ and $\langle -|\hat{\Pi}_{k}|-\rangle$, and
thus on $w_{k}$.
 If this bound is reached, then there can be only
one other element in the POM, which must be proportional to either $|+\rangle \langle +|$ or
$|-\rangle \langle -|$ to satisfy the completeness condition. Thus, if the 
optimal measurement strategy is
composed of more than three elements, then all of the elements must satisfy the
equation $\frac{\partial G}{\partial w_{k}} = 0$ as well as $\frac{\partial
G}{\partial \theta_{k}} = 0$.

It is possible to show that there are no values of the unknown multipliers
$\alpha_{1}$ and $\alpha_{2}$ which simultaneously satisfy $\frac{\partial
G}{\partial w_{k}} = 0$ for all four solutions of $\frac{\partial
G}{\partial \theta_{k}} = 0$, except in special cases where $F$ does not depend
on $\theta_{k}$ (for $\cos{\phi}_{k} = 1$).

Having established that the optimal strategy consists of three or less elements
we simplify the problem by applying the POM conditions (\ref{con1}) and (\ref{con2})
to obtain the most general three element mirror symmetric POM. This POM has only
one free parameter, $\cos{\Omega}$, and it is now a simple matter to maximise $F$.
The optimal measurement strategy is always found to be one or other of the two
mirror symmetric two element strategies.

Obtaining the optimal retransmission states is simply a matter of finding the
eigenvector of $\hat{O}_{k}$ corresponding to the larger of its two eigenvalues.
The determination of these states is detailed in appendix B.

\section{Complete Maximum Fidelity Strategy}

We summarise the results for the measurement-retransmission
strategy maximising
the fidelity for a set of three mirror symmetric states.

If
\begin{equation}
p( 1  - \cos{2\theta})[ p(1 - \cos{2\theta}) + \cos{2\theta}] = 0,
\label{bounder}
\end{equation}
then any POM consisting of elements of the form:
\begin{equation}
\hat{\Pi}_{k} = w_{k} \left( \begin{array}{cc}
1+\cos{\theta_{k}} & \sin{\theta_{k}} \\
\sin{\theta_{k}} & 1-\cos{\theta_{k}}
\end{array} \right);
\end{equation}
which satisfies the POM conditions (\ref{con1}, \ref{con2}, \ref{con3}) with
$\phi_{k} = 0$, will
maximise the fidelity.
 The optimal retransmission state for each element is then given by
\begin{equation}
|\phi_{k} \rangle = \left({Y_{k}}^{2} + 1\right)^{-\frac{1}{2}}(|+\rangle + Y_{k}|-\rangle),
\end{equation}
with $Y_{k}$ given by equation (\ref{Yk}).

If
\begin{equation}
p \ < \ -\frac{\cos{2\theta}}{1 - \cos{2\theta}}, \ p \ \neq \ 0, 
\end{equation}
then the unique optimal measurement strategy consists of the two elements:
\begin{eqnarray}
\hat{\Pi}_{0} & = & | + \rangle \langle + | \nonumber \\
\hat{\Pi}_{\pi} & = & | - \rangle \langle - |, \label{POMud}
\end{eqnarray}
and the optimal retransmission state is $| + \rangle$ if the result is $\hat{\Pi}_{0}$ and
$| - \rangle$ if the result is $\hat{\Pi}_{\pi}$.

If
\begin{equation}
p \ > \ -\frac{\cos{2\theta}}{1 - \cos{2\theta}}, \ p \ \neq \ 0,  \cos{2\theta} \
\neq \ 1, \label{peopt}
\end{equation}
then the unique optimal measurement strategy consists of the two elements:
\begin{eqnarray}
\hat{\Pi}_{+\frac{\pi}{2}} & = & \frac{1}{2}(| + \rangle + | - \rangle)( \langle
+ | + \langle - |) \nonumber \\
\hat{\Pi}_{-\frac{\pi}{2}} & = & \frac{1}{2}(| + \rangle - | - \rangle)( \langle
+ | - \langle - |),\label{POMlr}
\end{eqnarray}
with the optimal retransmission state for these elements given by
\begin{equation}
|\phi_{\pm \frac{\pi}{2}} \rangle = \left[1+(\sqrt{{\eta}^{2} + 1} - \eta)^{2}\right]^{-\frac{1}{2}}
\left[|+\rangle \pm (\sqrt{{\eta}^{2} + 1} - \eta)|-\rangle\right],
\label{rtmn}
\end{equation}
where $\eta$ is given by
\begin{displaymath}
\eta =  \frac{2p\cos{2\theta} + 1-2p}{2p\sin^{2}{2\theta}}.
\end{displaymath}

\section{Comments on the Optimal Strategy}

There are
only two
POMs, (\ref{POMud}) and (\ref{POMlr}), consisting of two pure state projector elements
 which are mirror
symmetric and confined to the plane
of the
original states.
At least one of these two POMs will always be the optimal measurement strategy.
The other will give least fidelity
of all POMs
composed of elements in that plane. In the case where these two give the same
fidelity, that is when
condition (\ref{bounder}) holds, any measurement strategy composed of elements in
that plane
will also be optimal. This condition (\ref{bounder}) can be decomposed into to two
obvious cases and one intriguing case. The first two are when there is only one
possible signal state ($p = 0$)
and when all of the signal states are identical ($\theta = 0$). The remaining case
corresponds to the situation that the sum of the density matrices of the states,
each multiplied by the probability of NOT selecting that state, sums to the
identity matrix, as described by the equation 
\begin{equation}
\sum_{j} (1-p_{j})|\psi_{j}\rangle\langle\psi_{j}| = \hat{1}.
\end{equation}
The meaning of this final condition remains unclear.

The retransmission states $|\phi_{k}\rangle$ for the $\theta_{k} = 0, \pi$ solution are simple
to understand as they are just the states that the POM elements project on to. The
origin of the retransmission states for the $\theta_{k} = \pm
\frac{\pi}{2}$ case
seems less transparent, but they are simply the states that the POM elements
project onto, rotated to increase their overlap with the a priori state of the signal.

We wish to compare the optimal measurement strategy for maximum fidelity with that for
minimum error. The $\theta_{k} = 0, \pi$ strategy is never optimal for
distinguishing between three mirror symmetric states
with least probability
of error.
The POM given by $\theta_{k} = \pm \frac{\pi}{2}$ is
the strategy which distinguishes
$|\psi_{3}\rangle$ and $|\psi_{2}\rangle$ (the mirror symmetric pair)
with minimum error probability.
This strategy is optimal for
distinguishing some mirror symmetric
sets of states with least error. These sets are given by the condition (\ref{pe2opt}). This
equation (\ref{pe2opt}) bears little resemblance to the
condition (\ref{peopt}) for the $\theta_{k} = \pm \frac{\pi}{2}$ strategy being
the optimal
fidelity measurement, save that both are satisfied for larger values of $p$.
This is to be expected as we
know from \cite{Bfid, F2Gen} that the optimal strategies must coincide for two equiprobable
states,
that is for $p=\frac{1}{2}$.
 The domains in which these
strategies are optimal, and the boundaries between them are shown
in figure 1.

\begin{figure}[ht]
\resizebox{\linewidth}{!}{\includegraphics*{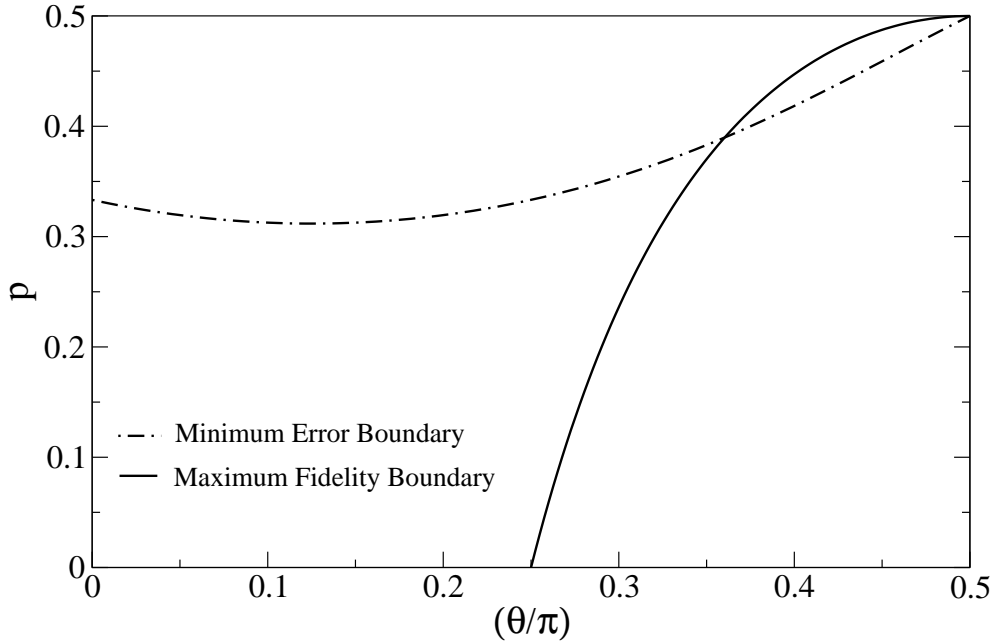}}
\label{fdomains}
\caption{A comparison of the domains in which the $\theta_{k} = \pm
\frac{\pi}{2}$ strategy is optimal for minimum error [equation
(\ref{pe2opt}), above the dash-dot line] and for maximum fidelity
[equation (\ref{peopt}), to the left of the solid line], in terms of
the state parameters $p$ and $\theta$.}
\end{figure}

From this diagram we can see that the region where $\theta_{k} = \pm \frac{\pi}
{2}$ is the fidelity
maximising measurement strategy (left of the solid line) contains most
of the region where this POM is
the minimum error measurement strategy (above the dot-dash line). There is, 
however, a sizable region
where the $\theta_{k} = \pm \frac{\pi}{2}$ strategy
maximises the fidelity without minimising the error probability and 
 a smaller
region where it minimises the error probability without maximising the fidelity.

It could be argued that the lack of correlation should not surprise us, as the
two boundaries have very different
significance.
The boundary (\ref{pe2opt}) for the minimum error strategy shows where the two
element (\ref{2elements}) and three
element (\ref{3elements}) strategies are identical in form, and the two element
 strategy is only optimal when one of
the three elements given by (\ref{3elements}) is not a positive operator.
Conversely the boundary for optimal
fidelity strategies corresponds to the situation where the two very different
strategies which are optimal
in their respective domains give the same fidelity. The fidelity is generally
maximised by a two element
strategy.

There is a unique strategy minimising the probability of error for all 
possible mirror symmetric sets of three qubit states, and a unique fidelity
maximising strategy everywhere not on the solid boundary in figure 1. Since
these unique strategies coincide only in the region of figure 1 to the left 
of the solid line and above the dot-dash line, it is clear 
 that the minimum
error strategy does not generally maximise the fidelity.

Two of the previous solutions \cite{Bfid, F2Gen} for a strategy maximising fidelity for a certain
class of state
sets overlap somewhat with the mirror symmetric states. The case of two
equiprobable pure
states clearly just corresponds to our mirror symmetric states with
$p = \frac{1}{2}$. Unsurprisingly
our solution is in complete agreement with \cite{Bfid, F2Gen} in predicting that
fidelity is
maximised by the minimum error measurement strategy in this case. The second
case is that
of the Trine states, an example of a set of symmetric states which is also dealt
 with in
\cite{Bfid}. Their solution predicts that any POM with elements in the common
plane of
the three Trine states will be optimal. It is easy to show that the Trine
states, with
$p = \frac{1}{3}$ and $\theta = 60^{o}$, satisfies the equality (\ref{eqcon}),
and hence also (\ref{bounder}),
and thus our solution
is also in accord here. It is interesting to note that we predict the same
solution for
a much wider class of state sets than just the Trine states, so it may be
possible to
extend the solution for the general set of symmetric states to a broader class of
sets.

Finally, in the course of the solution for a set of three mirror symmetric
states we only
referred to the nature and number of the states themselves in finding the
eigenvalues of
our $\hat{O}_{k}$ matrices. The only property of these eigenvalues that we
used in deriving
our solution was that the sum of them depended only on the values of
$\cos{\theta_{k}}$ and
${\cos}^{2}{\phi_{k}}$, and not otherwise on $\sin{\theta_{k}}$ or $\phi_{k}$.
The rest of the analysis was done using the coefficients of $\cos{\theta}_{k}$ ($A$, $B$ and
$C$ in the appended calculations) and holds true for any form of these coefficients. This
implies that for {\em any} set of qubit states for which this eigenvalue sum
has a similar dependence
only on $\cos{\theta_{k}}$ and ${\cos}^{2}{\phi_{k}}$, the optimal measurement
strategy will
again be either $\theta_{k} = \{\pm \frac{\pi}{2}\}$ or $\theta_{k} = \{0, \pi\}$,
or any set
of elements in the linear plane if the fidelity of these two strategies is
the same. In
particular this means our strategy maximising the fidelity for  a set of three
mirror
symmetric states is also the measurement strategy for maximising the fidelity
for
any number of mirror symmetric states sharing a common plane. The form of
the condition (\ref{eqcon})
will obviously become more complex in terms of the original variables when
there are
more states, but will be unchanged as a function of the coefficients of
$\cos{\theta_{k}}$ (\ref{eqcoeff}).

\section{Conclusions}

The fidelity of a measurement and reconstruction strategy is defined as the
average probability
that a subsequent measurement on the reconstructed state will identify it as
being identical
to the original state.
We sought to find a strategy which maximised the fidelity for a mirror symmetric
set of three
qubit states. This was done by parameterising the POM, evaluating the fidelity
as a
function of these parameters and conducting a variational calculation using
Lagrange's
method of undetermined multipliers to identify the sets of elements which
could constitute the maximum.

The optimal measurement strategy was found to be whichever of the two mirror symmetric
two element POMs
gives the larger fidelity for a given set of states,
and that any POM whose elements lie in the plane
of the states
is optimal if these two strategies give the same fidelity.
The optimal retransmission states were found using an eigenvector equation
(\ref{egvr}).

Unlike all previous solutions for maximum fidelity strategies, the minimum
error
measurement strategy does not generally maximise the fidelity of a set of
mirror symmetric states.

\ack

This work was supported by the UK Engineering and Physical Sciences Research Council,
the Marie Curie Fellowship Scheme of the European Commission, the Royal Society of
Edinburgh and the Scottish Executive Education and Lifelong Learning Department.

\appendix

\section{Derivation of the Fidelity Maximising Measurement}

In attempting to maximise the fidelity $F$ subject to the POM conditions
(\ref{con1}, \ref{con2}) it is helpful to begin by using Lagrange's method of
undetermined multipliers. We construct the function $G$:
\begin{equation}
G = F + \alpha_{1} \left(1 - \sum_{k} w_{k}\right) + \alpha_{2} \left(\sum_{k} w_{k}
\cos{\theta_{k}}\right).
\end{equation}

Varying $G$ with respect to $w_{k}$ gives us a restriction on the possible positions
 of the global maxima and minima of
$G$, which must be located either at a stationary point with respect to $w_{k}$:
\begin{equation}
\frac{\partial G}{\partial w_{k}} =  Q_{k}^{\frac{1}{2}} - \alpha_{1} +
\alpha_{2}\cos{\theta_{k}} = 0, \label{Gw}
\end{equation}
or at the maximum or minimum possible values of $w_{k}$:
\begin{equation}
w_{k:min} = 0, \ \ w_{k:max} = \frac{1}{1+|\cos{\theta}_{k}|};
\end{equation}
with this maximum arising from the fact that $w_{k}$ represents the combined
weight of the pair of POM elements corresponding to
 $\pm {\theta}_{k}$, and that the component of such a pair in
either the $|+\rangle$ or $|-\rangle$ direction cannot exceed 1. The minimum
value of $w_{k}$ clearly
corresponds to trivial
zero operators.

Since $Q_{k}$ is a function of $\cos{\theta_{k}}$ only, differentiating $G$ with
respect to $\theta_{k}$ gives
\begin{equation}
\frac{\partial G}{\partial \theta_{k}} = \frac{w_{k}}{2} Q_{k}^{-\frac{1}{2}}
\frac{\partial Q_{k}}{\partial \cos{\theta_{k}}} (-\sin{\theta_{k}})- \alpha_{2} w_{k}
\sin{\theta_{k}} = 0 \label{dtheta}
\end{equation}
at a stationary point of $G$. This has the nontrivial solutions:
\begin{equation}
\sin{\theta_{k}} = 0  \ => \ \theta_{k} = 0 \ or \ \pi \label{sine}
\end{equation}
and
\begin{equation}
\frac{\partial Q_{k}}{\partial \cos{\theta_{k}}} = -2\alpha_{2}Q_{k}^{\frac{1}{2}}, \label{alpha}
\end{equation}
as well as the $w_{k} = 0$ solution corresponding to the element not being part of the POM.

To simplify further analysis we
shall assign the coefficients of $\cos{\theta_{k}}$ in $Q_{k}$ to be
\begin{displaymath}
A = p\cos{2\theta} + \left(\frac{1}{2}-p\right),
\end{displaymath}
\begin{displaymath}
B = p\sin^{2}{2\theta} = \frac{1}{2} - C \ \geq \ 0,
\end{displaymath}
\begin{displaymath}
C = p\cos^{2}{2\theta} + \left(\frac{1}{2}-p\right) = \frac{1}{2} - B \ \geq \ 0;
\end{displaymath}
so that $Q_{k}$ is given by
\begin{displaymath}
Q_{k} = (A + C\cos{\theta_{k}})^{2} + B^{2} (1 - \cos^{2}{\theta_{k}}).
\end{displaymath}

We can then solve equation (\ref{alpha}) to find the remaining values of $\theta_{k}$
which satisfy $\frac{\partial G}{\partial \theta_{k}} = 0$. Since equation
(\ref{alpha}) contains only $\cos{\theta_{k}}$ terms, it is simplest to express the solution as
a value of $\cos{\theta_{k}}$ given in terms of our $A,B,C$ coefficients as
\begin{equation}
\cos{\theta_{k}} = \frac{1}{C^{2} - B^{2}}\left( -AC -
\alpha_{2}B\sqrt{\scriptstyle{\frac{A^{2} + B^{2} - C^{2}}{\alpha_{2}^{2} +
B^{2} - C^{2}}}} \right), \label{cosine}
\end{equation}
which can only take one value for a given POM since $\alpha_{2}$ must have a
single value for all of the elements
of one POM.

We can now say that any measurement maximising the fidelity has at most four
elements, corresponding
to the four solutions of $\frac{\partial G}{\partial \theta_{k}} = 0$
(\ref{dtheta}) for $\theta_{k}$,
given by equations (\ref{sine}) and (\ref{cosine}). For each of these
possible elements, either
equation (\ref{Gw}) holds (a stationary point of $G$ with respect to $w_{k}$)
or $w_{k}$
takes its maximum or minimum value.

Now we must consider whether it is possible to have all four of these elements
present in one
POM, i.e. that none of the weight factors $w_{k}$ are zero for these elements.
Clearly this
implies that none of them take their maximum values either, since the maximum
value of
$w_{k}$ for an element (or mirror symmetric pair of elements) is found by noting
 that
the positivity and completeness of the POM implies that neither the
$|+\rangle \langle +|$ or $|-\rangle \langle -|$ component of any element
(or pair) can exceed 1.
If any element saturates this bound, there can be only {\em one} more nonzero
element
corresponding to either $|+\rangle \langle +|$ or $|-\rangle \langle -|$ to satisfy
 the completeness condition (\ref{eq:POMcond2}).

Since no weight factor $w_{k}$ can attain its maximum value when there are
four nonzero
elements present in the POM, the equation (\ref{Gw}) must be simultaneously
satisfied
for all four elements. This occurs if there is a pair of values for $\alpha_{1}$
 and
$\alpha_{2}$ which will satisfy (\ref{Gw}) for all three solutions for
$\cos{\theta_{k}}$ obtained from (\ref{dtheta}).

For $\theta_{k} = 0$ and $\pi$, equation (\ref{Gw}) gives
\begin{equation}
\theta_{k} = 0: |A+C| = \alpha_{1} - \alpha_{2}, \label{th0}
\end{equation}
\begin{equation}
\theta_{k} = \pi: |A-C| = \alpha_{1} + \alpha_{2}, \label{thpi}
\end{equation}
which fixes both $\alpha_{1}$ and $\alpha_{2}$ for any measurement strategy
containing both of these elements.
These values must satisfy the equation (\ref{Gw}) for the value of
$\cos{\theta_{k}}$ given
by (\ref{cosine}).
Since the multipliers $\alpha_{1}$ and $\alpha_{2}$ can only take the values
$\pm A, \mp C$ respectively or $\pm C, \mp A$ respectively for any values of
$A$ and $C$ when (\ref{th0}, \ref{thpi}) hold, it is simple to show that
equation (\ref{Gw}) can only be satisfied for this value of $\cos{\theta_{k}}$
when
either
\begin{equation}
A^{2} + B^{2} - C^{2} = 0 \ \mathrm{ or} \ B = 0. \label{cases}
\end{equation}
Which of these two conditions is relevant depends on the relative magnitudes and
signs of $A$ and $C$.
Thus we see that we can only have a four element
POM in certain special cases.

Examining these special cases shows that in each of them $Q_{k}$ is the square
of some linear
function of $\cos{\theta_{k}}$ which is either positive for all $\theta_{k}$ or
negative for all $\theta_{k}$.
If $|Q_{k}^{\frac{1}{2}}|$ is any linear function of $\cos{\theta_{k}}$, it
can be
shown by application of the POM conditions (\ref{con1},
\ref{con2}) that $F$ does not depend on any
$\cos{\theta_{k}}$
and thus the fidelity is  constant for any measurement strategy composed of
elements confined to the plane of the states (that is for which
$\cos{\phi_{k}} =  1$).

For the general case where $F$ does depend on the strategy chosen we now know
that there
is no strategy composed of four or more elements which can be a maximum or minimum of
the fidelity.
Denoting the solution of equation (\ref{cosine}) in the range $0 \ \leq \
\theta_{k} \ \leq \ \pi $ as $\theta_{k} = \Omega $, there are two
possibilities for three element strategies: case (i)
\begin{equation}
\theta_{k} = \pi, \pm \Omega; \ 1 \ \geq \ \cos{\Omega} \ \geq \ 0;
\end{equation}
or case (ii)
\begin{equation}
\theta_{k} = 0, \pm \Omega; \ -1 \ \leq \ \cos{\Omega} \ \leq \ 0.
\end{equation}

Both of the two element measurement strategies which are mirror symmetric in
this basis are special cases of these three element  strategies, and are located
at the edge of the domains of the three element strategies. The
$\theta_{k} = \{0, \pi \}$ strategy corresponds to $\cos^{2}{\Omega} = 1$ in either
of the above cases,
and the $\theta_{k} = \{\pm \frac{\pi}{2}\}$ strategy corresponds to
$\cos{\Omega} = 0$.

For these three element strategies the POM conditions (\ref{con1}) and
(\ref{con2}) place a strict limit on  the values
of the weight factors $w_{k}$. Denoting the weights of the
$\theta_{k} = 0, \pi$
elements and the $\theta_{k} = \pm \Omega$ pair by $w_{0}$, $w_{\pi}$ and
$w_{\Omega}$ respectively,
we have either:

(i) for $\cos{\Omega} \ \geq \ 0$
\begin{equation}
w_{\pi} + w_{\Omega} = 1, \ w_{\pi} - w_{\Omega}\cos{\Omega} = 0
\end{equation}
which gives
\begin{equation}
w_{\Omega} = \frac{1}{1 + \cos{\Omega}}, \ w_{\pi} = \frac{\cos{\Omega}}{1 + \cos{\Omega}};
\end{equation}

or (ii) for $\cos{\Omega} \ \leq \ 0$
\begin{equation}
w_{0} + w_{\Omega} = 1, \  w_{0} + w_{\Omega}\cos{\Omega} = 0
\end{equation}
which gives
\begin{equation}
w_{\Omega} = \frac{1}{1 - \cos{\Omega}}, \ w_{0} = \frac{- \cos{\Omega}}{1 - \cos{\Omega}}.
\end{equation}

Now we need simply differentiate $F$ with respect to $\cos{\Omega}$ for each of these two strategies
and select the largest value of $F$ from any stationary points and the two limiting two element
strategies. Both of these strategies automatically satisfy all of the POM conditions so we
no longer need to use Lagrange's method of undetermined multipliers.

Case (i):

For the $\theta_{k} = \pi, \pm \Omega$ case the stationarity equation
$\frac{\partial F}{\partial \cos{\Omega}} = 0$ can be rearranged and squared to
obtain
\begin{equation}
B^{2}(C^{2} - B^{2} - A^{2})(1 + \cos{\Omega})^{2} = 0.
\end{equation}
The only solutions to this equation are the aforesaid special cases
(\ref{cases}) and $\cos{\Omega} = -1$,
which is not allowed since $\cos{\Omega} \ \geq \ 0$ in this case. We conclude that
there
are no stationary points of $F$ for this set of strategies and the maximum and minimum
of
the fidelity for these strategies must correspond to the two element strategies which
define the end points of
our variation (i.e. $\cos{\Omega} = 0$ or $1$).

Case (ii):

Similarly for the $\theta_{k} = 0, \pm \Omega$ case, $\frac{\partial F}{\partial \cos{\Omega}} = 0
$ implies that
\begin{equation}
B^{2}(C^{2} - B^{2} - A^{2})(1 - \cos{\Omega})^{2} = 0.
\end{equation}
As before, the only solutions to this are our two special cases (\ref{cases})
and the single value of
$\cos{\Omega} = 1$, which is not in the domain for this strategy. We can thus conclude
that there are
no stationary points of $F$ for either strategy and our global maximum and
minimum must correspond
to the $\theta_{k} = \{\pm \frac{\pi}{2}\}$ and $\theta_{k} = \{0, \pi\}$ strategies
which are at the end points of both of our three element strategy domains.

The fidelity for each of the two strategies which must constitute our maximum and
minimum are

for $\theta_{k} = \{\pm \frac{\pi}{2}\}$
\begin{equation}
F = \frac{1}{2} + \sqrt{B^{2} + A^{2}},\label{fidlr}
\end{equation}

and for $\theta_{k} = \{0, \pi\}$
\begin{equation}
F = \frac{1}{2} + \frac{|A+C|}{2} + \frac{|A-C|}{2},\label{fidud}
\end{equation}
which is just equal to a half plus the larger of $|A|$ or $|C|$.

The larger of these two fidelities, (\ref{fidlr}) and (\ref{fidud}), will be the
maximum fidelity, and the corresponding POM will be
the optimal measurement strategy. For the $\theta_{k} = \{0, \pi\}$ strategy
to be optimal, we must have $C \ \geq \ |A|$ since the fidelity of
the $\theta_{k} = \{\pm \frac{\pi}{2}\}$ strategy is always at least
$\frac{1}{2} + |A|$.
Thus the $\theta_{k} = \{0, \pi\}$
strategy is only uniquely optimal if
\begin{equation}
A^{2} + B^{2} - C^{2} \ < \ 0, \label{eqcoeff}
\end{equation}
which can be restated in terms of the original variables $p$ and $\theta$ as
\begin{equation}
p \ < \ -\frac{\cos{2\theta}}{1 - \cos{2\theta}}.
\end{equation}

The two strategies give the same fidelity when the relevant condition in
(\ref{cases})
holds. This corresponds to the special case where any POM consisting of elements
confined to the plane of the states is optimal. These conditions  can be written
in terms of $p$ and $\theta$ as:
\begin{equation}
p( 1  - \cos{2\theta})[ p(1 - \cos{2\theta}) + \cos{2\theta}] = 0,
\end{equation}
 that is either
\begin{equation}
p = 0, \label{onestate}
\end{equation}
or
\begin{equation}
\cos{2\theta}=1, \label{samestate}
\end{equation}
or
\begin{equation}
p = -\frac{\cos{2\theta}}{1 - \cos{2\theta}}. \label{eqcon}
\end{equation}
The meaning of two of these three cases is clear: equation (\ref{onestate}) is the
case where there is only one possible signal state and equation (\ref{samestate})
describes the case where all three states are identical. The fidelity obviously
cannot depend on the measurement strategy at all in these cases. The third of
these cases, equation (\ref{eqcon}), is less obvious. In fact it corresponds to the
identity:
\begin{equation}
\sum_{j} (1-p_{j})|\psi_{j}\rangle\langle\psi_{j}| = \hat{1}, \label{ident}
\end{equation}
that is when the sum of the density operators of the states normalised to the
prior probability of NOT selecting that state is the identity operator.

\section{Retransmission States}

In equation (\ref{eigen}) we found the optimal retransmission states to be
the eigenvectors of
$\hat{O}_{k}$, which is given by equation (\ref{Ok}). This determines the best
retransmission
state for
any measurement we choose to make, not only the optimal measurement. The
optimal retransmission state $|\phi_{k}\rangle$ depends on the possible states
of the original signal
\{$|\psi_{j}\rangle$\} and on the direction of the {\em corresponding} measurement operator
$\hat{\Pi}_{k}$. 
Since $|\phi_{k}\rangle$ does not depend on the weight ($w_{k}$) of this element
or on the rest of the POM, $|\phi_{k}\rangle$ will be the optimal retransmission
state for any POM containing an element in this direction. This is useful as we
need only
find $|\phi_{k}\rangle$ for each possible element, without having to consider
the strategy in which the element occurs.
It could therefore be said that the optimal retransmission
state $|\phi_{k}\rangle$ depends on the result of the measurement (given by
the direction of 
$\hat{\Pi}_{k}$) 
rather than
on the measurement strategy
(that is the experiment whose outcome was $k$).

It is simplest to find the states \{$|\phi_{k}\rangle$\} if we consider the
folowing three cases separately: $\theta_{k} = 0$, $0 < |\theta_{k}| < \pi$ and
$\theta_{k} = \pi$.

For $\theta_{k} = 0$ our $\hat{O}_{k}$ matrix has eigenvectors $|+\rangle$ and
$|-\rangle$. The larger eigenvalue
belongs to $|+\rangle$ if $A + C > 0$. Since we found previously that we must
have $C \ \geq \ |A|$ for $\theta_{k} = 0, \pi $ to be the best strategy, it
 is always the case that the optimal retransmission state for the element
$\Pi_{0} $ is $|+\rangle$ when we have employed the optimal measurement
 strategy.

Similarly, for $\theta_{k} = \pi$  our $\hat{O}_{k}$ matrix again has eigenvectors
$|+\rangle$ and
$|-\rangle$. The larger eigenvalue now
belongs to $|-\rangle$ if $A - C < 0$, which is again always true when the
optimal strategy includes a  $\theta_{k} = \pi$ element.

For any other value of $\theta_{k}$, such as $\theta_{k} = \pm \frac{\pi}{2}$,
 we must find the general solution for the
 eigenvector corresponding to the larger eigenvalue of a $2 \times 2$ real
 Hermitian
 matrix. We need only study nondiagonal matrices since the $\hat{O}_{k}$ matrix
 is diagonal when $\theta_{k} \neq 0$ or $\pi$ only for trivial sets of states
 \{$|\psi_{j}\rangle, p_{j}$\}. The equation for the unnormalised eigenvectors
is
\begin{equation}
\left( \begin{array}{cc}
R_{k} & S_{k} \\
S_{k} & P_{k}
\end{array}\right)
\left( \begin{array}{c}
1 \\
Y_{k\pm}
\end{array}\right)
 = \nu_{k\pm}
\left( \begin{array}{c}
1 \\
Y_{k\pm}
\end{array}\right)  \label{egvr}
\end{equation}
 which gives a value for $Y_{k\pm}$ in terms of the matrix elements:
\begin{equation}
Y_{k\pm} = \frac{P_{k} - R_{k}}{2S_{k}} \pm \frac{\sqrt{(P_{k} - R_{k})^2 +
4S^2_{k}}}{2S_{k}}, \label{Yk}
\end{equation}
where the $\pm$ in $Y_{k\pm}$ in this equation corresponds to the two eigenvalues
$\nu_{k\pm}$, so the eigenvector of interest is that which contains $Y_{k+}$.

From the general form of $\hat{O}_{k}$ given in equation (\ref{odef}) we can
identify the elements of our eigenvector equation as
\begin{eqnarray}
R_{k} & = & 2p\cos^{2}{\theta}(1+\cos{2\theta}\cos{\theta_{k}}) +
(1-2p)(1+\cos{\theta_{k}}) \nonumber \\
P_{k} & = & 2p\sin^{2}{\theta}(1+\cos{2\theta}\cos{\theta_{k}}) \nonumber \\
S_{k} & = & p\sin^{2}{2\theta}\sin{\theta_{k}},
\end{eqnarray}
from which we can identify $Y_{k+}$ and thus find and normalise
$|\phi_{k}\rangle$. It can be readily appreciated that the form of these states
is not simple.

For the case where $\theta_{k} = \pm \frac{\pi}{2}$, we identify the parameter
$\eta$ as
\begin{equation}
\eta = \frac{R_{k} - P_{k}}{2|S_{k}|} =  \frac{2p\cos{2\theta}
 + (1-2p)}{2p\sin^{2}{2\theta}}, \label{eta}
\end{equation}
then $Y_{\pm \frac{\pi}{2}}$ is given by
\begin{equation}
Y_{\pm \frac{\pi}{2}} = \pm  \left( \sqrt{{\eta}^{2} + 1} - \eta \right).
\label{yak}
\end{equation}
The optimal retransmission state for this strategy is given by
\begin{equation}
|\phi_{\pm \frac{\pi}{2}} \rangle = \frac{1}{\sqrt{1+(Y_{\pm \frac{\pi}{2}})^{2}}}
[|+\rangle + (Y_{\pm \frac{\pi}{2}})|-\rangle]. \label{retlr}
\end{equation}
It is clear from the form of $Y_{\pm \frac{\pi}{2}}$ (\ref{yak}) that the
retransmission state given by
(\ref{retlr})
is, as expected, on
the same side of the Bloch sphere as the corresponding POM element $\hat{\Pi}_{\pm
\frac{\pi}{2}}$. Further
analysis of the physical meaning of these states is possible by rewriting the
$\eta$ parameter as
\begin{equation}
\eta = \frac{\langle + | \hat{\rho}_{T} | + \rangle - \langle - | \hat{\rho}_{T} |
- \rangle}{2p\sin^{2}{2\theta}},
\end{equation}
where $\hat{\rho}_{T}$ is the state Bob assigns to the signal before making his
measurement and is given by
\begin{equation}
\hat{\rho}_{T} = \sum_{j} p_{j} | \psi_{j} \rangle \langle \psi_{j} |.
\end{equation}

It is clear that $\hat{\rho}_{T}$ is also a measure of the `average state' of the
signal sent by Alice and that $\eta$ is positive if the $| + \rangle \langle + |$
component of $\hat{\rho}_{T}$ is larger than the $| - \rangle \langle - |$
component (the `average state' is closer to $| +
\rangle$ than $| - \rangle$) and negative if the converse is true.
Furthermore we see that $|Y_{\pm \frac{\pi}{2}}|$ is larger than one if $\eta$
is negative and smaller than one is $\eta$ is positive, so that the
retransmission states $|\phi_{\pm \frac{\pi}{2}} \rangle$ are shifted from $| +
\rangle \pm | - \rangle$ towards the `average state', $\hat{\rho}_{T}$.

\end{document}